\documentclass[twocolumn,,mathlines,left,switch]{aastex701}

\usepackage{booktabs}
\usepackage{graphicx}
\usepackage{amsmath} 
\usepackage{multirow}
\usepackage{bm}
\usepackage{comment}

\begin{document}

\title{Quasi-Keplerian Be-Star Disks with Mimicking Viscosity}
\author{Michel Cur\'e}
\affiliation{Instituto de F\'isica y Astronom\'ia, Universidad de Valpara\'iso, Av. Gran Breta\~na 1111, Valpara\'iso, Chile}
\email[show]{michel.cure@uv.cl}

\author{Ignacio Araya}
\affiliation{Centro Multidisciplinario de F\'isica, Vicerrector\'ia de Investigaci\'on, Universidad Mayor, 8580745 Santiago, Chile}
\email[]{ignacio.araya@umayor.cl}

\author{Rodrigo Meneses}
\affiliation{Escuela de Ingenier\'ia Civil, Universidad de Valpara\'iso, General Cruz 222, Valpara\'iso, Chile}
\email[]{odrigo.meneses@uv.cl}

\author{Mat\'ias Montesinos}
\affiliation{Departamento de F\'isica, Universidad T\'ecnica Federico Santa Mar\'ia, Avenida Espa\~na 1680, Valpara\'iso, Chile}
\email[]{matias.montesinosa@usm.cl}

\author{Roberto O. J. Venero}
\affiliation{Departamento de Espectroscop\'ia, Facultad de Ciencias Astron\'omicas y Geof\'isicas, Universidad Nacional de La Plata, Paseo del Bosque S/N, B1900FWA La Plata, Buenos Aires, Argentina}
\affiliation{Instituto de Astrof\'isica de La Plata, CCT La Plata, CONICET--UNLP, Paseo del Bosque S/N, B1900FWA La Plata, Argentina}
\altaffiliation{Member of the Carrera del Investigador Cient\'ifico, CONICET, Argentina}
\email[]{roberto@fcaglp.unlp.edu.ar}

\author{Catalina Arcos}
\affiliation{Instituto de F\'isica y Astronom\'ia, Universidad de Valpara\'iso, Av. Gran Breta\~na 1111,  Valpara\'iso, Chile}
\email[]{catalina.arcos@uv.cl}

\author{Abigali Rodriguez}
\affiliation{Instituto de F\'isica y Astronom\'ia, Universidad de Valpara\'iso, Av. Gran Breta\~na 1111, Valpara\'iso, Chile}
\affiliation{Departamento de Astronom\'ia, Universidad de Chile, Casilla 36-D, Santiago, Chile}
\email[]{arodrigu@das.uchile.cl}

\author{Lydia S. Cidale}
\affiliation{Departamento de Espectroscop\'ia, Facultad de Ciencias Astron\'omicas y Geof\'isicas, Universidad Nacional de La Plata, Paseo del Bosque S/N, B1900FWA La Plata, Buenos Aires, Argentina}
\affiliation{Instituto de Astrof\'isica de La Plata, CCT La Plata, CONICET--UNLP, Paseo del Bosque S/N, B1900FWA La Plata, Argentina}
\altaffiliation{Member of the Carrera del Investigador Cient\'ifico, CONICET, Argentina}
\email[]{lydia@fcaglp.fcaglp.unlp.edu.ar}

\begin{abstract} 
Classical Be stars are fast-rotating B-type stars with gaseous quasi-Keplerian disks formed by equatorial ejection of material. While the viscous decretion disk (VDD) model reproduces many observed properties, the role of radiative line driving in shaping these disks remains unclear.
We investigated the combined influence of viscosity and radiative acceleration on the hydrodynamic structure of Be star disks by coupling the m-CAK theory of line-driven winds with a mimicking viscous prescription governed by the parameter $\gamma_{\rm vis}$.
We solved the steady-state hydrodynamic equation of motion using \textsc{Hydwind} for typical B-type stellar parameters in transonic $\Omega$-slow outflows. We analyzed the velocity and density structures and derived the mass-loss rates and radial velocities at the adopted outer integration radius, $r=50\,R_\ast$.
The combined action of line driving and viscosity yields regular m-CAK $\Omega$-slow solutions for equatorial outflow with a VDD-inspired rotational prescription. For quasi-Keplerian exponents ($\gamma_{\rm vis} \simeq 0.5$) and near-critical rotation ($\Omega \approx 0.96$--$0.99$), the models produce an outflowing disk with an m-CAK-type critical point at $r_{\rm c} \lesssim 20$--$30\,R_\ast$. At $50\,R_\ast$, these quasi-Keplerian solutions reach radial velocities of $76.9$--$139.2\,\mathrm{km\,s^{-1}}$. Within the present 1D parameterized framework, the m-CAK line force yields stationary solutions without imposing an outer boundary condition.
Our results provide a controlled 1D test of how a VDD-inspired rotational prescription modifies the topology of stationary m-CAK $\Omega$-slow solutions. The model is an exploratory bridge toward future non-Sobolev multidimensional radiation-hydrodynamic treatments that recover quasi-Keplerian rotation and modest outflow velocities of Be disks.
\end{abstract}
    
\section{Introduction}
Classical Be stars (CBes) are fast-rotating main-sequence B-type stars that form an equatorial gas-rotating disk around their equators. The inner part of the disk, usually $\lesssim$ 20 stellar radii, is geometrically thin and optically thick and rotates in a quasi-Keplerian orbit \citep{Quirrenbach1997,Meilland2012,Rivinius2013}. These disks, referred to as decretion disks, are built from mass ejected from the equatorial stellar surface, which acquires sufficient velocity and angular momentum to orbit the star. Once the material is ejected from the star, its subsequent evolution is generally governed by gravity, viscous forces, and radiative acceleration \citep[for the latest review, see][and references therein]{rivinius2026}.

Currently, the best theoretical framework describing the evolution of these disks once formed is the viscous decretion disk (VDD) model proposed by \citet{lee1991}. In this model, viscosity redistributes the angular momentum of the circumstellar material and, under the assumption of Keplerian gas motion, results in a steady, thermally stable structure. In this model, the viscosity strength is parameterized according to the prescription of \citet{Shakura1973}. 
Applying the VDD model to investigate the disk density evolution in Be stars,  \citet{Haubois2012} determined that only a small fraction (approximately 1\%) of the ejected mass attains sufficient angular momentum to migrate outward, whereas the majority falls back onto the star. 
Further exploring disk build-up and dissipation within the VDD framework, \citet{Rimulo2018} investigated 81 outburst events in 54 Classical Be stars (CBes) by employing non-local thermodynamic equilibrium radiative transfer models to reproduce light curves. Their study revealed that the viscosity parameter was higher during the disk formation phase than during the dissipation phase, indicating that disk formation occurred more rapidly than disk dissipation.

\citet{okazaki2001} solved the hydrodynamic equations for the VDD model, assuming isothermal conditions and a radiative force modeled with an ensemble of optically thin lines using an ad hoc model from \citet{chen1994}. They investigated the transition between near-Keplerian and angular momentum conservation, finding transonic solutions where the sonic point\footnote{The critical (or singular) point in the works of \citet{okazaki2001} and \citet{cure2022} corresponds to the sonic point.} is located more than 100 stellar radii from the stellar surface. The flow was highly subsonic and nearly Keplerian within this region, whereas the angular momentum was conserved outside this region. \citet{okazaki2001} demonstrated that the topology of the sonic point is highly sensitive to viscosity, yielding nodal solutions at higher viscosities and saddle solutions at lower viscosities. In that formulation, the sonic radius and the specific angular momentum at the sonic point are eigenvalues of the coupled radial- and angular-momentum equations.

\citet{cure2022} revisited the viscous transonic decretion disk model from Okazaki with methodological improvements. They introduced a new solution scheme in which the eigenvalue is radial rather than angular, demonstrating that viscosity can collapse different rotational velocity profiles into a nearly unique solution, near the stellar surface. However, \citet{okazaki2001} and \citet{cure2022} used a radiative-force model that relied on the work of \citet{chen1994}, which is based on an ensemble of optically thin lines. This phenomenological description, while helpful in demonstrating the existence of nodal and saddle transonic solutions, lacks the physical rigor of radiation-hydrodynamic approaches. Their results showed that the location of the sonic point can vary from $\sim 50$ to over 600 stellar radii, depending on the line force parameters, underscoring the sensitivity of the solutions to the line-force prescription.

In contrast, the present work adopts the standard m-CAK (modified Castor, Abbott \& Klein) formalism, which provides a Sobolev-based treatment of line-driven acceleration in stellar winds and includes the finite-disk correction factor. For the viscous treatment, we adopted the parameterized prescription of \citet{dearaujo1995}, in which a viscosity-mimicking parameter controls the deviation from angular momentum conservation. Thus, while \citet{okazaki2001} and \citet{cure2022} employed a rigorous viscosity treatment with a simplified radiation force, we followed the complementary path of using a standard Sobolev-based line-force prescription with a parameterized viscosity, allowing us to assess the relative importance of these two mechanisms.
This study aims to investigate the combined influence of radiative line driving and viscosity on the hydrodynamic structure of Be-star disks by coupling the m-CAK theory with a viscosity-mimicking prescription governed by the exponent $\gamma_{\rm vis}$.
In this sense, our model should be viewed as an intermediate step between purely viscous decretion disks and more complete non-Sobolev, multidimensional simulations of Be star outflowing disks.

It is important to clarify the distinction between the m-CAK critical point ($r_{\rm c}$) and the gas sonic point ($r_{\rm s}$). In a Parker-type thermal wind, the critical and sonic points coincide. In the m-CAK formalism, however, the line force depends explicitly on the velocity gradient; consequently, the $\Omega$-slow critical point is located away from the stellar surface and lies within the supersonic flow regime. Therefore, we interpret the innermost subsonic region as a boundary or launching layer and restrict the physical interpretation
of the m-CAK line force to regions where $v>a$, where $a$ is the sound speed. For more details about the topological analysis of the m-CAK $\Omega$-slow solution, see \citet{cure2004}.

In addition to viscosity, radiative ablation may systematically remove material from the inner parts of Be disks via radiative acceleration, producing a disk--wind structure above and below the equatorial plane \citep{Kee2019}. In the context of outer-disk evolution, \citet{KOM2011} studied the link between angular momentum loss and mass-loss rates, showing that processes such as binarity, thermal expansion, and radiative ablation influence the outer disk and should be incorporated into stellar evolution codes. Similarly, \citet{Kurfurst2014} studied large disks and their dependence on viscosity and temperature, although both studies neglected radiation forces. Observationally, \citet{Klement17} emphasized the importance of extending VDD studies to the outer disk by compiling ultraviolet-to-radio spectral energy distributions (SEDs). They found a turn-down between the far-infrared and radio regimes in five of six Be stars, which was reproduced with truncated disks of $26$--$108\,R_{\ast}$, likely owing to binary tidal forces.

Non-radial line forces and disk ablation have been investigated using sophisticated radiative transfer techniques, ranging from the analytic work of \citet{Gayley2001} to the multidimensional simulations of \citet{Kee2016}. Non-Sobolev line-driving studies, such as those of \citet{Gomez2003}, further highlight the complex time-dependent structure that arises when the full radiation field is treated explicitly.

Recent time-dependent simulations of rotating m-CAK winds have shown that the transition between fast and slow wind regimes can remain continuous across the $(\Omega,\delta)$ parameter space, including regions where
stationary methods fail to recover the solutions \citep{Montesinos2026,Melina2026}. These calculations also revealed complex supersonic velocity structures and the possible coexistence of multiple hydrodynamic branches, indicating that the absence of a stationary solution does not necessarily imply 
physical instability.

Motivated by these considerations, the present formulation, which couples the m-CAK line-force prescription with a viscosity-mimicking rotational law in a controlled 1D framework, is hereafter referred to as the mimicking-VDD model (hereafter m-VDD).
The remainder of this paper is organized as follows: Section 2 reviews the m-VDD model and introduces our procedure for solving the equations of motion.  Section 3 presents the numerical results of the $\Omega$-slow solution. Finally, Sections 4 and 5 present the discussion and conclusions, respectively.
	
\section{The mimicking viscous line-driven hydrodynamic model}
\label{sec2}

The equation of motion (EoM) for radiation-driven stellar winds was developed within the framework of the m-CAK theory. The formulation is based on the conservation of mass and momentum for a one-component isothermal fluid in a stationary, spherically symmetric regime, which includes the effects of rotation and neglects heat conduction and magnetic fields \cite[see, e.g.,][]{lamers1999,cure2023}. The continuity equation is as follows:
\begin{equation}
\label{eq:mass}
4\pi \,r^{2}\,\rho \,v\,=\,\dot{M}\, ,
\end{equation}where $v=v(r)$ denotes the radial component of the wind velocity, $\rho$ is the mass density, and $\dot{M}$ represents the stellar mass-loss rate. On the other hand, the momentum conservation equation is expressed as follows:
\begin{equation}
\label{eq:mEoMlo}
    v(r)\frac{d v(r)}{d r}=-\frac{1}{\rho}\frac{d P}{d r}-\frac{GM(1-\Gamma)}{r^2}+g^{\rm line}+\frac{v_{\phi}^2(r)}{r}\,,
\end{equation}where $M$ is the stellar mass, $P$ denotes the fluid pressure, $\Gamma$ the radiative acceleration caused by Thomson scattering in terms of the gravitational acceleration, $v_{\phi}$ the rotational velocity component of the wind in the equatorial plane, and $g^{\rm line}$ is the acceleration due to an ensemble of spectral lines \citep{abbott1982}.

In this study, the viscosity is modeled through the rotational centrifugal term, following \citet{dearaujo1995}, as
\begin{equation}
    v_{\phi}(r)=v_{\phi}(R_{\ast})\, \left(\frac{R_{\ast}}{r}\right)^{\gamma_{\mathrm{vis}}}\, ,
    \label{eq:2d}
\end{equation}where $R_{\ast}$ is the stellar radius. The disk kinematics are governed by the parameter $\gamma_{\rm vis}$ (with $0.5 \lesssim \gamma_{\rm vis} < 1$), which {\it{mimics}} the effect of the viscosity. In this framework, $\gamma_{\rm vis}$ acts as a control parameter that allows the description of different configurations: for $\gamma_{\rm vis}=1$, angular momentum is conserved, while for $\gamma_{\rm vis}\simeq 0.5$ a quasi-Keplerian disk is attained. 
The surface rotational velocity $v_{\phi}(R_{\ast})$, in Eq. ~(\ref{eq:2d}) is given by 
\begin{equation}
\label{eq:vcrit}
    v_{\phi}(R_{\ast})=\Omega \, v_{\rm crit}\,,
\end{equation}where 
\begin{equation}
\label{ecu-vel}
v_{\rm crit}=\sqrt{GM(1-\Gamma)/R_{\ast}} \,\, ,
\end{equation}is the critical rotational speed, and $\Omega$ is the dimensionless rotation parameter.

To express Eq.~(\ref{eq:mEoMlo}) solely in terms of the radial velocity \( v(r) \), we use the normalized m-CAK line-force formulation given by \citet{cure2004}, based on the finite-disk extensions of the CAK theory developed by \citet{friend1986} and \citet{ppk1986}. The pressure term is described by an ideal gas, $P = a^{2}\rho$, along with Eq.~\eqref{eq:mass}, where $a$ denotes the isothermal sound speed. The line acceleration is
\begin{equation}
\label{eq:gline}
    g^{\rm line}=\dfrac{C}{r^{2}}\, \left( \dfrac{n_{\rm E}}{W(r)} \right)^{\delta} CF\left(r,v,\frac{dv}{dr}\right)\, \left(r^{2}\,v\, \dfrac{dv}{dr}\right)^{\alpha} \, ,
\end{equation}here, $C$ is the eigenvalue of the problem,
\begin{equation}\label{eq:eigenvalue}
    C=\Gamma GM\:k \left(\frac{4\pi}{\sigma_{\rm e}\:v_{\rm th}\dot{M}}\right)^{\alpha},
\end{equation}where $\sigma_{\rm e}$ is the electron-scattering opacity per unit mass and $v_{\rm th}$ is the proton thermal speed. The finite-disk correction factor is \citep[see, e.g.,][]{friend1986,ppk1986,cure2004}
\begin{equation}
CF = \frac{(1+\sigma)^{1+\alpha}-(1+\sigma \mu_*^2)^{1+\alpha}}{\sigma (1 + \alpha)(1+\sigma)^{\alpha}(1-\mu_*^2)}\, ,
\label{ecu-coefi3}
\end{equation}where
\begin{equation}
\sigma =  \frac{r}{v} \frac{\partial v}{\partial r} - 1\, ,
\label{ecu-sigmav}
\end{equation}and $\mu_*^2=1-(R_{\ast}/r)^2$. The geometrical dilution factor is $W(r)=(1/2)\,[1-\sqrt{1-R_{\ast}^2/r^2}]$. The quantities $k$, $\alpha$, and $\delta$ are the standard line-force parameters, and $n_{\rm E}$ is the electron density in units of $10^{11}\,\mathrm{cm}^{-3}$; the ionization parameter $\delta$ and this electron number density normalization follow \citet{abbott1982}.

Equation~(\ref{eq:gline}) computes the m-CAK line acceleration using the Sobolev approximation. 
In this approximation, the Doppler shift produced by the accelerating flow localizes the interaction between the radiation field and a given spectral line to a narrow resonance region. Consequently, the line optical depth is determined locally by the velocity gradient through the Sobolev optical depth. The finite disk correction factor accounts for the angular extension of the stellar radiation field. Because the present calculations are restricted to stationary hydrodynamic solutions integrated from a prescribed low-velocity photospheric boundary, the application of Eq.~(\ref{eq:gline}) to the innermost subsonic layers should not be interpreted as a full non-local radiative transfer treatment of the wind base.
Rather, Eq.~(\ref{eq:gline}) is used as an approximate Sobolev prescription for the line acceleration, allowing the solution to connect the lower boundary to the transonic transition and supersonic m-CAK winds. This does not invalidate the m-CAK solution at low velocities; it only indicates that the detailed radiative transfer in the subsonic launching region is not resolved by the proposed formulation. This interpretation is consistent with the non-Sobolev pure-absorption study of \citet{Poe1990}, who analyzed stationary radiatively driven winds and found that the steady problem near the sonic point has a nodal topology with two possible positive slopes and does not yield a unique global stationary solution in the non-Sobolev formulation.

Hence, substituting and rearranging terms in Eq.~\eqref{eq:mEoMlo}, the 1D equation of motion for the equatorial m-VDD wind model reads as follows
\begin{eqnarray}
0 &=& \left(1 - \frac{a^2}{v^2}\right) r^{2} v \frac{dv}{dr} + GM(1-\Gamma) \left(1 - \Omega^2 \left(\frac{R_{\ast}}{r}\right)^{\bm{{{2\gamma_{\mathrm{vis}} - 1}}}}\right) \nonumber \\
  & & - 2a^2 r - C  \, CF\, \left(r^{2} v \frac{dv}{dr}\right)^{\alpha} \, \left(\frac{n_{\rm E}}{W(r)}\right)^{\delta} \,,
\label{eq:araujo1}
\end{eqnarray}subject to the boundary condition 
\begin{equation}
\label{eq:densup}
\rho(R_{\ast})=\rho_{\ast}    \,,
\end{equation}where $\rho_{\ast}$ is the density at the stellar surface \citep[see details in][]{cure2004}. In Eq~(\ref{eq:araujo1}), we have highlighted in boldface the exponent ${{2\gamma_{\mathrm{vis}} - 1}}$, which is the only difference from the EoM in the study of \citet{cure2004}. 

\section{Results}
In this study, we present numerical solutions to hydrodynamic equations that describe radiation-driven stellar winds. Our main goal was to obtain $\Omega$-slow solutions using this viscosity-mimicking model, which was computed using the {\sc{Hydwind}} code \citep{cure2004}. This code provides the corresponding velocity and density profiles, as well as the resulting mass-loss rates, for the equatorial plane of fast-rotating Be stars. The {\sc{Hydwind}} code solves the EoM using the independent variable $u = -R_{\ast}/r$.

The adopted stellar parameters correspond to those of a B1~V classical Be star. It is well established that radiative acceleration is mainly produced by optically thin and weak lines \citep[see][]{okazaki2001} in these systems, leading to a value of the m-CAK line-force parameter $\alpha$ that is smaller than that usually assumed for fast-wind regimes. For the present models, we adopted $\alpha = 0.41$, considering that when $\alpha<0.5$, thin lines contribute more than thick lines. Additionally, a standard value of $k=0.32$ was employed for the m-CAK parameter, while the outflowing disk wind was maintained in a state of frozen ionization, indicated by $\delta = 0$. We adopted a solar helium abundance by number, $Y_{\mathrm{He}} = N(\mathrm{He})/N(\mathrm{H})=0.1$.

Motivated by viscous transonic calculations that predict highly subsonic radial drift in the inner disk \citep{okazaki2001} and by interferometric studies that favor quasi-Keplerian disk rotation \citep{Quirrenbach1997,Meilland2012}, we used $\gamma_{\mathrm{vis}} = 0.51$ and $0.52$ to reproduce this dynamical behavior, in which the azimuthal velocity scales as $v_{\phi} \propto r^{-\gamma_{\mathrm{vis}}}$. For comparison, we also computed solutions with $\gamma_{\mathrm{vis}} = 1.0$, which corresponds to angular momentum conservation. The numerical search for such solutions is challenging because the {\sc{Hydwind}} code requires an appropriate trial function to initiate the integration. In addition, time-dependent calculations have shown that apparent gaps in the stationary m-CAK solution space can result from the limitations of stationary numerical methods and that the recovered solution may depend on the initial flow configuration \citep{Montesinos2026}. In practice, convergent solutions were obtained for rotation rates of $\Omega = 0.96$ and $\Omega = 0.99$. By using such high values of $\Omega$, we ensured that the models remained in the regime of $\Omega$-slow solutions. All parameters are summarized in Tab.~\ref{tab:stellar_parameters}.

For $\gamma_{\mathrm{vis}}=0.51$ and $0.52$, the adopted rotational prescription maintains nearly Keplerian orbital motion while allowing radial expansion. These specific hydrodynamic and interferometric results \citep[][and references therein]{Rivinius2013,rivinius2026} support the use of $\gamma_{\mathrm{vis}}\simeq0.5$ to describe the kinematic regime explored in this study.

\begin{figure*}
    \centering
    \includegraphics[scale=0.3]{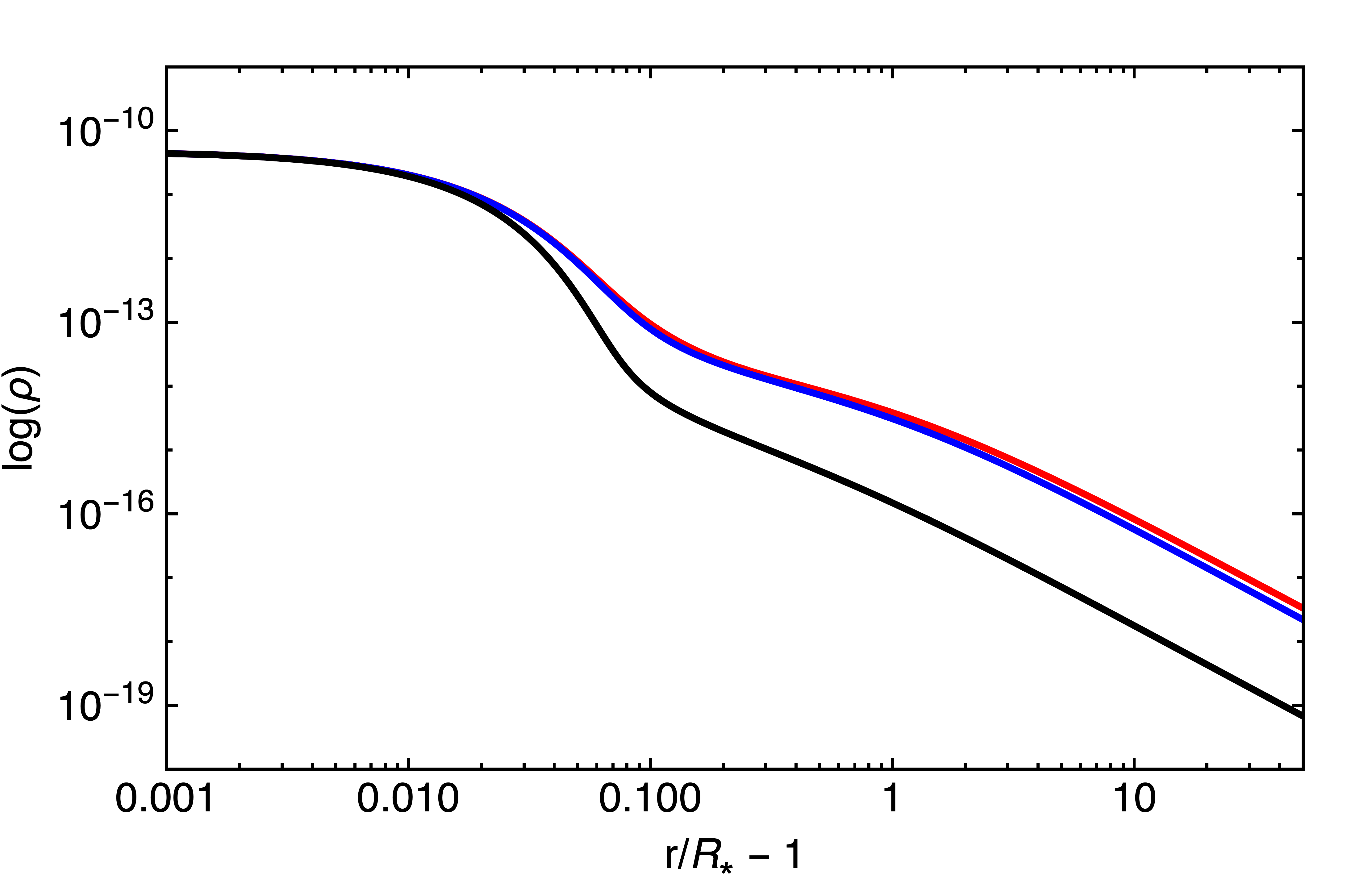}    \includegraphics[scale=0.3]{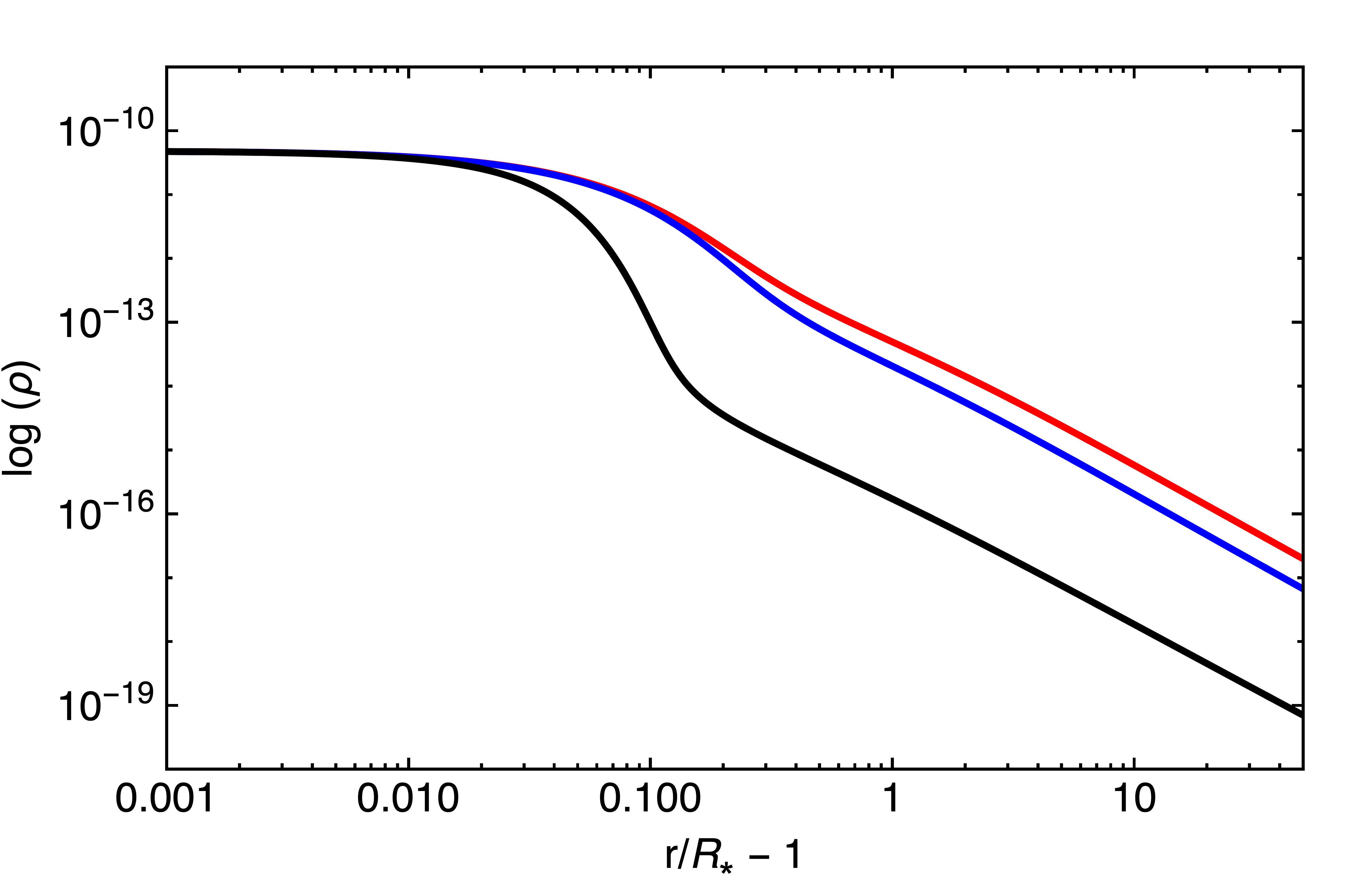}\\
    \includegraphics[scale=0.3]{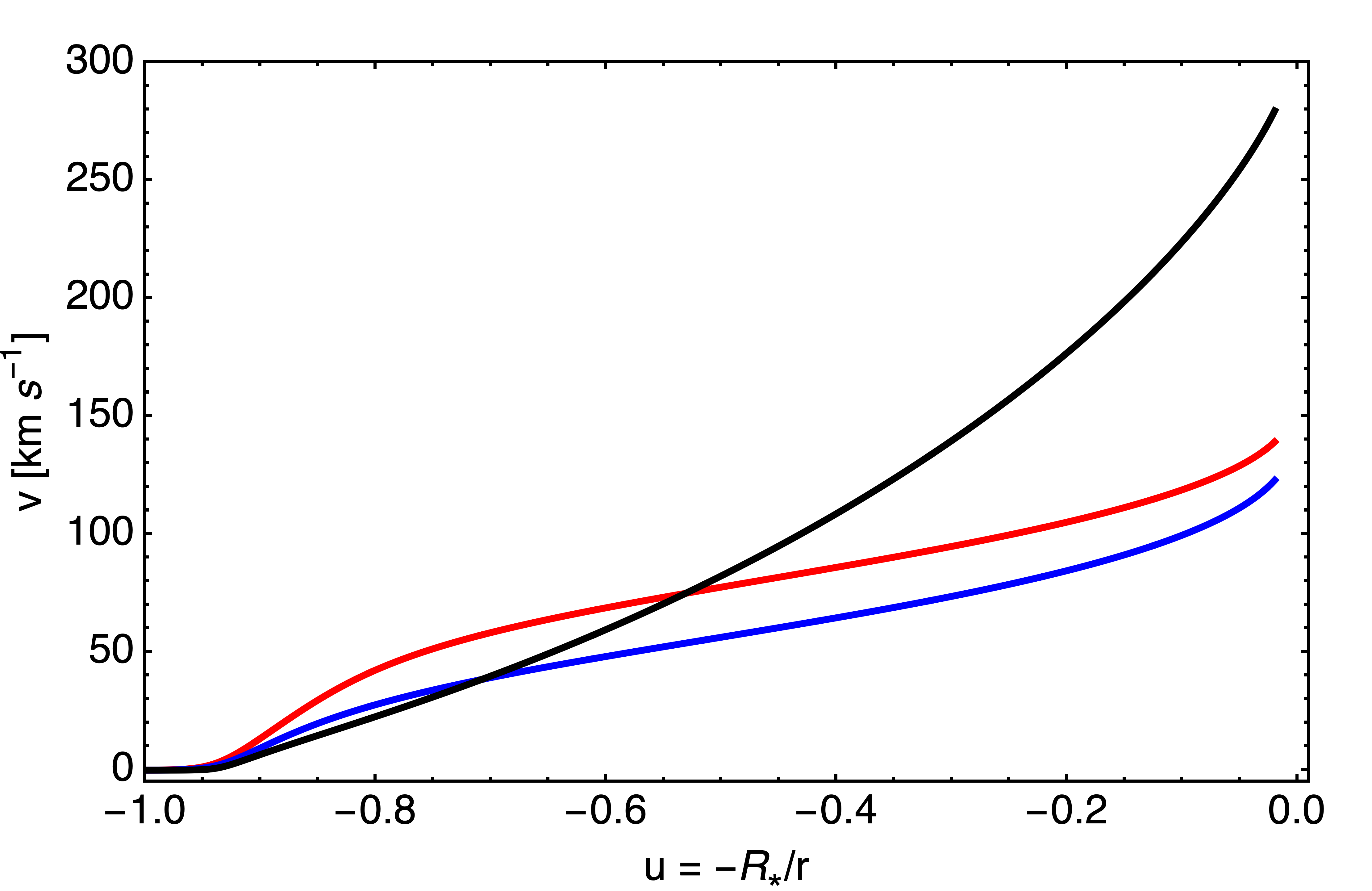}    
    \includegraphics[scale=0.3]{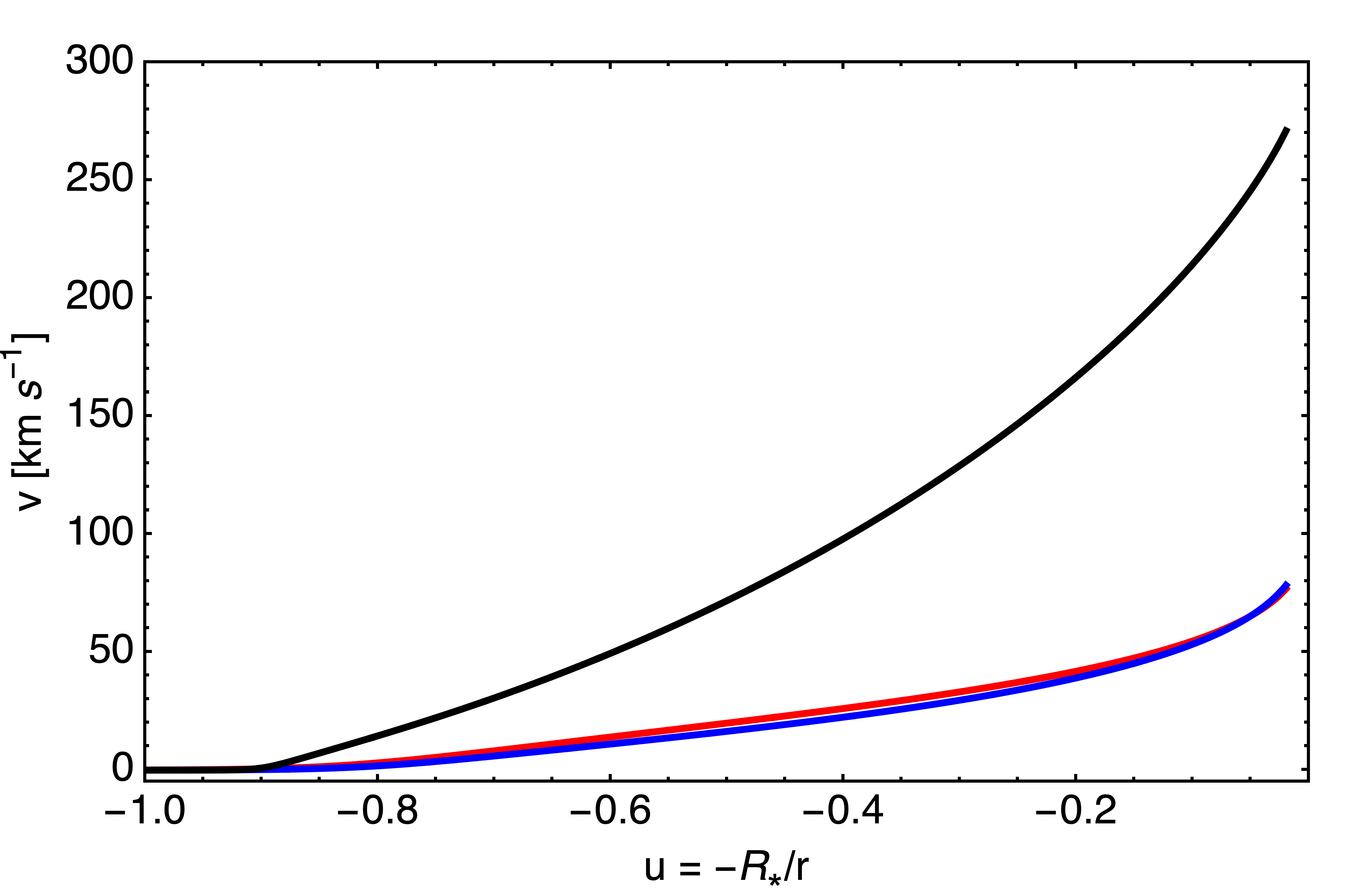}
    \caption{
    m-VDD models. Three distinct models for the $\gamma_{\mathrm{vis}}$ exponent are presented: $\gamma_{\mathrm{vis}}=0.51$ (red solid line), $\gamma_{\mathrm{vis}}=0.52$ (blue solid line), and $\gamma_{\mathrm{vis}}=1.0$ (black solid line).
    Logarithmic density profiles, $\log(\rho)$ versus $r/R_{\ast}-1$, are shown for the selected star rotating at $\Omega=0.96$ (left panels) and $\Omega=0.99$ (right panels). 
    The lower panels illustrate the corresponding velocity profiles $v(u)$ as a function of $u=-R_{\ast}/r$.
    The line force parameters and stellar properties were fixed (Tab.~\ref{tab:stellar_parameters}).
    }
    \label{fig:denplusvel}
\end{figure*}

Figure~\ref{fig:denplusvel} shows the logarithmic density profiles (upper panels) and velocity profiles (lower panels) for $\Omega = 0.96$ (left column) and $\Omega = 0.99$ (right column). Table~\ref{tab:results} summarizes the corresponding mass-loss rates, radial velocities evaluated at $r=50\,R_{\ast}$, and m-CAK critical-point radii.

The density profiles exhibited similar global behaviors for all cases. However, the model with $\Omega = 0.99$ presented higher densities owing to its lower expansion velocities (according to the continuity equation, Eq.~\ref{eq:mass}). In contrast, the $\Omega = 0.96$ case showed an earlier density decline associated with a noticeable {\it bump} in the velocity profile (see the bottom left panel of Fig.~\ref{fig:denplusvel}).

The velocity profiles exhibited a stronger dependence on the viscous-mimicking parameter, $\gamma_{\mathrm{vis}}$. For $\gamma_{\mathrm{vis}} = 1.0$, both rotation rates ($\Omega = 0.96$ and $\Omega = 0.99$) yielded similar (density and velocity) profiles. In contrast, for both considered $\gamma_{\mathrm{vis}}$ quasi-Keplerian values, the curves differ significantly. In particular, for $\Omega = 0.96$, $v(u)$ exhibits a pronounced {\it bump} in the range $-0.85 \lesssim u \lesssim -0.5$, particularly for $\gamma_{\mathrm{vis}} = 0.51$. For $\Omega = 0.99$, the overall shape of the profiles maintained the same structure as that in the $\gamma_{\mathrm{vis}} = 1.0$ case.

An additional interesting feature is observed in the lower panels of Fig.~\ref{fig:denplusvel}, where the velocity profile corresponding to $\gamma_{\mathrm{vis}} = 0.52$ remains below that of $\gamma_{\mathrm{vis}} = 0.51$ over most of the integration domain, except near $u \lesssim -0.05$, where both curves intersect. This behavior suggests that the smallest radial velocity at the adopted outer boundary is not attained precisely at $\gamma_{\mathrm{vis}}=0.5$, but at a marginally higher value.
Concerning the results shown in Tab.~\ref{tab:results}, a slight change in $\gamma_{\rm vis} \gtrsim 0.5$ produces a significant change in the mass-loss rate, and the effect becomes stronger as $\Omega \to 1$. For the quasi-Keplerian cases ($\gamma_{\mathrm{vis}}=0.51$ and $0.52$), $v(50R_\ast)$ spans $76.9$--$139.2\,\mathrm{km\,s^{-1}}$. The angular-momentum-conserving models ($\gamma_{\mathrm{vis}}=1$) reach larger values of $271.1$--$279.6\,\mathrm{km\,s^{-1}}$. These are model velocities at the finite outer integration radius and are not assumed to be asymptotic terminal velocities.

\begin{table}[h]
\centering
\caption{Stellar and line-force parameters adopted in this work.}
\label{tab:stellar_parameters}
\begin{tabular}{lc}
\hline\hline
Parameter & Value \\
\hline
$\mathrm{SpT}$ & B1~V \\
$T_{\mathrm{eff}}$ [K]     & 21000 \\
$R_{*}$ [R$_{\odot}$]      & 4.5 \\
$\log g$                   & 4.0 \\
$\alpha$                        & 0.41\\  
$k$                             & 0.32\\  
$\delta$                        & 0.0\\  
$\Omega$                        & 0.96, 0.99\\
$\gamma_{\mathrm{vis}}$  & 0.51, 0.52, 1.0 \\
\hline
\end{tabular}
\end{table}

\begin{table*}[ht]
\centering
\caption{Wind parameters and critical-point radii obtained for the B1~V model. The velocities $v_{50}\equiv v(50R_{\ast})$ are evaluated at the adopted outer integration radius and should not be interpreted as asymptotic terminal velocities.}
\label{tab:results}
\small
\setlength{\tabcolsep}{4pt}
\begin{tabular}{c| ccc |ccc}
\hline
\hline
\multirow{2}{*}{$\gamma_{\mathrm{vis}}$} & \multicolumn{3}{c|}{$\Omega = 0.96$} & \multicolumn{3}{c}{$\Omega = 0.99$} \\ \cline{2-7}
 & $\dot{M}$ & $v_{50}$ & $r_{\rm c}/R_\ast$ & $\dot{M}$ & $v_{50}$ & $r_{\rm c}/R_\ast$ \\
 & [$10^{-9}\,\mathrm{M_{\odot}\,yr^{-1}}$] & [km s$^{-1}$] & & [$10^{-9}\,\mathrm{M_{\odot}\,yr^{-1}}$] & [km s$^{-1}$] & \\ \hline
0.51 & 2.38 & 139.2 & 10.73 & 7.83 & 76.9 & 11.48 \\
0.52 & 1.39 & 122.9 & 19.25 & 2.70 & 78.4 & 20.99\\
1.00 & 0.098 & 279.6 & 23.74 & 0.098 & 271.1 & 24.31 \\ \hline
\end{tabular}
\end{table*}

\section{Discussion}
The results obtained in this study indicate that a mimicking-viscosity prescription, combined with the m-CAK radiative line force, provides a controlled parameterized framework for exploring steady, equatorial, line-driven outflow solutions with VDD-inspired rotation. This approach allows us to investigate the interplay between radiative driving and viscous effects in the regime that connects line-driven outflows to viscous disk dynamics. We stress that the use of a Sobolev-based m-CAK line force in the inner, subsonic parts of the disk should be interpreted as a parameterization of radiative driving, not as an exact description of the line force. Because the equations are not vertically integrated disk equations and do not include vertical density stratification, continuum optical depth effects, or non-radial radiative fluxes, the present solutions should not be interpreted as complete VDD solutions. A fully consistent treatment would require non-Sobolev, multidimensional radiative transfer, and non-radial line forces, as explored, for example, by \citet{Gayley2001,Kee2016,Gomez2003}.

Our results are consistent with the general picture established by \citet{okazaki2001}, who showed that transonic viscous decretion solutions exist for a wide range of viscosity parameters in the Shakura \& Sunyaev model \citep{Shakura1973}, with the sonic point located beyond $\sim 100\,R_\ast$. In his treatment, the line acceleration was modeled through an ensemble of optically thin lines, while the sonic radius and the specific angular momentum at the sonic point were the two eigenvalues of the coupled radial- and angular-momentum equations. The structure of those steady isothermal solutions is independent of the mass-decretion rate \citep{okazaki2001}. The solutions presented here instead adopt the m-CAK formalism for line-driven acceleration, providing a physically grounded description of the radiative force while treating viscosity in a simplified parameterized form that mimics angular-momentum redistribution. In this sense, our approach complements that of \citet{cure2022}, who reformulated the viscous transonic problem using a radial eigenvalue rather than the angular eigenvalue adopted by \citet{okazaki2001}.

A notable outcome is that the Sobolev-based m-CAK line-force prescription yields $\Omega$-slow-type solutions with a sonic point near the stellar surface and the critical (singular) radii listed explicitly in Table~\ref{tab:results}. For the quasi-Keplerian prescriptions ($\gamma_{\rm vis}=0.51$ and $0.52$), these critical radii fall within approximately $20$--$30\,R_\ast$. The m-CAK critical radius is a hydrodynamic singular point, not an outer disk radius. Nevertheless, its radial scale overlaps the $26$--$108\,R_\ast$ truncation radii inferred by \citet{Klement17} from ultraviolet-to-radio SED modeling. Establishing any physical connection between these two scales requires a self-consistent disk radiation-hydrodynamic treatment.

Within the present parameterized 1D framework, the results suggest that radiative acceleration can modify slow quasi-Keplerian outflow solutions when coupled to a viscosity-mimicking rotational prescription.

Our model may also have implications for angular-momentum loss from rapidly rotating B stars. The computed mass flux provides a basis for a future quantitative estimate of the associated angular-momentum transport; such an estimate is beyond the present 1D formulation. In this context, the m-VDD solutions can serve as a hydrodynamic reference for future non-Sobolev simulations of Be disks, helping to isolate the effects of line-driving prescriptions and viscosity on the outflowing disk structure.

\section{Conclusions}

In this study, we introduced the m-VDD model, a controlled 1D hydrodynamic formulation for exploring viscosity-mimicking, line-driven equatorial outflow solutions around rapidly rotating B stars. The combined prescription produces transonic $\Omega$-slow solutions with slow radial expansion and either quasi-Keplerian rotation ($\gamma_{\rm vis}=0.51$ and $0.52$) or angular-momentum-conserving rotation ($\gamma_{\rm vis}=1$), within the limitations of a 1D Sobolev-based formulation. At the adopted outer integration radius of $50\,R_\ast$, the quasi-Keplerian models reach radial velocities of $76.9$--$139.2\,\mathrm{km\,s^{-1}}$.

A complete topological analysis is still required to characterize the velocity profiles and identify all physically meaningful solution branches. This step was not performed because the present model is an intermediate stage toward a more comprehensive description. As discussed by \citet{cure2022}, a more realistic treatment should combine a Shakura--Sunyaev viscosity prescription \citep{Shakura1973} with an accurate line-force formalism. The present work follows the complementary approach of coupling a detailed m-CAK radiative force to the simplified viscosity-mimicking prescription of \citet{dearaujo1995}.

Future developments should combine the Shakura--Sunyaev-like viscosity prescription with m-CAK hydrodynamics, time-dependent evolution, and radiative transfer, while including non-isothermal and multidimensional effects. A detailed topological analysis will then be required to investigate the full set of steady and transonic solutions at high rotation rates. These extensions are essential for understanding the observed variability of Be disks and their interactions with magnetic fields or binary companions in the angular-momentum evolution of rapidly rotating massive stars.

\begin{acknowledgments}
MC \& CA acknowledge the support from Centro de Astrof\'isica de Valpara\'iso.
MC, CA \& IA thank the support from ANID FONDECYT projects 1230131 and 1261498. MM and MC acknowledge the financial support of the Fondecyt regular project 1241818. AR acknowledges funding from ANID-Subdirecci\'on de Capital Humano/Doctorado Nacional/2022-21221841. This project was funded by the European Union (Project 101183150 - OCEANS). This work has been made possible by AWS-U.Chile-NLHPC credits. Powered@NLHPC: This research was partially supported by the NLHPC's supercomputing infrastructure (ECM-02).  ROJV and LSC acknowledge financial support from Universidad Nacional de La Plata (Programa de Incentivos 11/G192 and 11/G193).
\end{acknowledgments}
	
\bibliography{cites}{}
\bibliographystyle{aasjournalv7}

\end{document}